\documentclass[aps,prd,twocolumn,twoside,superscriptaddress,floatfix,nofootinbib,hyperref]{revtex4-1}
\usepackage{epsfig}
\usepackage{epstopdf}
\input{epsf.sty}
\usepackage{epsf}
\usepackage{color}
\usepackage{bm}
\newcommand{\ba}{\begin{eqnarray}}
\newcommand{\ea}{\end{eqnarray}}
\newcommand{\be}{\begin{equation}}
\newcommand{\ee}{\end{equation}}

\newcommand\lsim{\mathrel{\rlap{\lower4pt\hbox{\hskip1pt$\sim$}}
        \raise1pt\hbox{$<$}}}
\newcommand\gsim{\mathrel{\rlap{\lower4pt\hbox{\hskip1pt$\sim$}}
        \raise1pt\hbox{$>$}}}

\newcommand{\jcap}{{J.~Cosm.~Astrop.~Phys.}}

\newcommand{\aap}{{Astron.~Astrophys.}}
\newcommand{\apjl}{{Astrophys.~J.~Lett.}}
\newcommand{\apjs}{{Astrophys.~J.~Supp.}}

\newcommand{\mnras}{{Mon.~Not.~R.~Astron.~Soc.}}

\input{epsf}

\usepackage{graphicx}
\usepackage[large]{subfigure}
\usepackage{amssymb, amsmath}
\usepackage[amssymb]{SIunits}
\usepackage{hyperref}
\usepackage{url}
\usepackage{aas_macros}
\usepackage{natbib}
\begin{document}
%%%%%%%%%%%%%%
\title{Cosmological Constraints from Moments of the Thermal Sunyaev-Zel'dovich Effect}
\author{J.~Colin Hill}
\affiliation{Department of Astrophysical Sciences, Princeton
  University, Princeton, NJ 08544 \\
  jch@astro.princeton.edu}
\author{Blake D.~Sherwin}
\affiliation{Department of Physics, Princeton University, Princeton, NJ 08544\\
bsherwin@princeton.edu}
\date{\today}

\begin{abstract}
In this paper, we explain how moments of the thermal Sunyaev-Zel'dovich (tSZ) effect can constrain both cosmological parameters and the astrophysics of the intracluster medium (ICM).  As the tSZ signal is strongly non-Gaussian, higher moments of tSZ maps contain useful information.  We first calculate the dependence of the tSZ moments on cosmological parameters, finding that higher moments scale more steeply with $\sigma_8$ and are sourced by more massive galaxy clusters.  Taking advantage of the different dependence of the variance and skewness on cosmological and astrophysical parameters, we construct a statistic, $|\langle T^3 \rangle|/\langle T^2 \rangle^{1.4}$, which cancels much of the dependence on cosmology (i.e., $\sigma_8$) yet remains sensitive to the astrophysics of intracluster gas (in particular, to the gas fraction in low-mass clusters).  Constraining the ICM astrophysics using this statistic could break the well-known degeneracy between cosmology and gas physics in tSZ measurements, allowing for tight constraints on cosmological parameters.  Although detailed simulations will be needed to fully characterize the accuracy of this technique, we provide a first application to data from the Atacama Cosmology Telescope and the South Pole Telescope.  We estimate that a \emph{Planck}-like full-sky tSZ map could achieve a $\lesssim 1$\% constraint on $\sigma_8$ and a $1\sigma$ error on the sum of the neutrino masses that is comparable to the existing lower bound from oscillation measurements.
\end{abstract}

\maketitle
\section{Introduction}
\label{sec:intro}
The thermal Sunyaev-Zel'dovich (tSZ) effect is a spectral distortion of the cosmic microwave background (CMB) caused by inverse Compton scattering of CMB photons off hot electrons in the intracluster medium (ICM) of galaxy clusters~\cite{Sunyaev-Zeldovich1970}.  The tSZ effect, which is largest on arcminute angular scales, has traditionally been studied either through observations of individual clusters --- both pointed measurements~\cite{Reeseetal2011,Plaggeetal2012, Lancasteretal2011,AMIetal2011} and recent detections in blind surveys~\cite{Marriageetal2011, Williamsonetal2011,Plancketal2011} --- or indirectly through its contribution to the CMB power spectrum~\cite{Dunkleyetal2011,Reichardtetal2011}.  %In recent years many clusters have been detected blindly in mm-wave surveys using the Atacama Cosmology Telescope (ACT) \jch{add citations} and the South Pole Telescope \jch{add citations}.  The \emph{Planck} satellite has also released a preliminary catalog of tSZ clusters \jch{add citation} and will soon release a full-sky catalog.  While studying individual clusters through their tSZ signal can provide a wealth of information about the ICM astrophysics of each object, 

The goal of many direct studies is to characterize the masses and redshifts of a well-defined sample of clusters, and thereby reconstruct the high-mass end of the halo mass function, an important cosmological quantity that is sensitive to a number of parameters.  These parameters include $\sigma_8$, the rms fluctuation amplitude on scales of $8 \, h^{-1}$ Mpc. However, one can also constrain these parameters through their influence on the tSZ power spectrum, which is exceptionally sensitive to $\sigma_8$ in particular~\cite{Komatsu-Seljak2002}.  In recent years, many studies have obtained competitive constraints on $\sigma_8$ from the power spectrum~\cite{Dunkleyetal2011,Reichardtetal2011}.  This approach is advantageous not only because the tSZ signal is extremely sensitive to $\sigma_8$, but also because it does not require the measurement of individual cluster masses, which is a difficult procedure.  Unfortunately, it does require a precise understanding of the pressure profile of the ICM gas for clusters over a wide range of masses and redshifts.  Consequently, systematic errors due to theoretical uncertainty in the astrophysics of the ICM have remained comparable to or greater than statistical errors in tSZ power spectrum measurements~\cite{Dunkleyetal2011,Reichardtetal2011}.  This situation has hindered the progress of tSZ measurements in providing cosmological constraints.

In this paper, we propose a method to reduce the theoretical systematic uncertainty in tSZ-derived cosmological constraints by combining different moments of the tSZ effect.  Although previous studies have investigated tSZ statistics beyond the power spectrum~\cite{Coorayetal2000,Rubino-Martin-Sunyaev2003, Holderetal2007,Munshietal2011}, none have attempted to find a tSZ observable that isolates the dependence on either astrophysical or cosmological parameters.  In $\S$II, we compute the variance (second moment) and skewness (third moment) of the tSZ signal for a variety of ICM astrophysics scenarios while varying $\sigma_8$, which allows us to probe the mass and redshift dependence of each statistic, as well as to derive their dependence on cosmology. We find that the skewness is not only more sensitive to $\sigma_8$ than the variance, but is also dominated by contributions from higher-mass clusters, for which the ICM astrophysics is better-constrained by existing observations.  In $\S$III, we use these results to find a particular combination of the variance and skewness --- the ``rescaled skewness'' --- that only depends weakly on the underlying cosmology while remaining sensitive to the ICM astrophysics prescription.  We test this statistic using a variety of models for the ICM, which should span the space of reasonable theoretical possibilities.  In $\S$IV, we apply this method to data from the Atacama Cosmology Telescope (ACT)~\cite{Fowleretal2007,Swetzetal2011} and the South Pole Telescope (SPT)~\cite{Carlstrometal2011,Schafferetal2011} to demonstrate its feasibility and to obtain a first weak constraint on the ICM astrophysics.  In $\S$V, we estimate the extent to which this approach can increase the precision of future constraints on $\sigma_8$ and the sum of the neutrino masses through tSZ measurements.  We also highlight applications to other parameters that affect the tSZ signal through the mass function.

We assume a concordance $\Lambda$CDM cosmology throughout, with parameters taking their maximum-likelihood WMAP5 values (WMAP+BAO+SN)~\cite{Komatsuetal2009} unless otherwise specified.  All masses are quoted in units of $M_{\odot}/h$, where $h \equiv H_0/(100 \, \mathrm{km} \, \mathrm{s}^{-1} \, \mathrm{Mpc}^{-1})$ and $H_0$ is the Hubble parameter today.

\section{Calculating tSZ Moments}
\label{sec:Npt}
\subsection{Background}
The tSZ effect leads to a frequency-dependent change in the observed
CMB temperature in the direction of a galaxy group or cluster.  Neglecting relativistic
corrections (which are relevant only for the most massive systems)~\cite{Nozawaetal2006},
the temperature change $T$ at angular position ${\bm \theta}$ with respect to the center of a
cluster of mass $M$ at redshift $z$ is given by~\cite{Sunyaev-Zeldovich1970}
\ba
\label{eq.tSZdef}
\frac{T({\bm \theta}, M, z)}{T_{\mathrm{CMB}}} & = & g_{\nu} y({\bm \theta}, M, z) \\
 &= & g_{\nu} \frac{\sigma_T}{m_e c^2} \int_{\mathrm{LOS}} P_e \left( \sqrt{l^2 + d_A^2 |{\bm \theta}|^2}, M, z \right) dl \,, \nonumber
\ea
where the tSZ spectral function $g_{\nu} = x\,\mathrm{coth}(x/2)-4$ with $x \equiv h\nu/k_B T_{\mathrm{CMB}}$,
$y$ is the Compton-$y$ parameter, $\sigma_T$ is the Thomson scattering cross-section, $m_e$ is the electron
mass, and $P_e({\bm r})$ is the ICM electron pressure at location ${\bm r}$ with respect to the cluster center.  In this work we only consider spherically symmetric pressure profiles with $P_e({\bm r}) = P_e(r)$. Also, we calculate all observables at $\nu = 150$ GHz, where the tSZ effect is observed as a decrement in the CMB temperature in the direction of a cluster (i.e., $g_{\nu} < 0$).  Note that the integral in Eq.~(\ref{eq.tSZdef}) is taken along the line of sight such that $r^2 = l^2 + d_A(z)^2 \theta^2$, where $d_A(z)$ is the angular diameter distance to redshift $z$ and $\theta \equiv |{\bm \theta}|$ is the angular distance from the cluster center in the plane of the sky.  Given a spherically symmetric pressure profile, Eq.~(\ref{eq.tSZdef}) implies that the temperature decrement profile is azimuthally symmetric in the plane of the sky, that is, $T({\bm \theta},M,z) = T(\theta,M,z)$.  Finally, note that the electron pressure is related to the thermal gas pressure via $P_{th} = P_e (5 X_H+3)/2(X_H+1) = 1.932 P_e$, where $X_H=0.76$ is the primordial hydrogen mass fraction.

In order to calculate moments of the tSZ effect, we assume
that the distribution of clusters on the sky can be adequately described by
a Poisson distribution (and that contributions due to clustering are negligible,
which is valid on sub-degree scales)~\cite{Cole-Kaiser1988,Komatsu-Kitayama1999}.
The $N^{\mathrm{th}}$ moment at zero lag is then given by
\be
\label{eq.Npoint}
\langle T^N \rangle = \int dz \frac{dV}{dz} \int dM \frac{dn(M,z)}{dM} \int d^2 \theta \, T (\theta,M,z)^N \, ,
\ee
where $dV/dz$ is the comoving volume per steradian at redshift $z$, $dn(M,z)/dM$ is the comoving number density of halos of mass $M$ at redshift $z$ (the halo mass function), and $T(\theta,M,z)$ is given by Eq.~(\ref{eq.tSZdef}).  Our fiducial integration limits are $0.005 < z < 4$ and
$5 \times 10^{11} M_{\odot}/h < M < 5 \times 10^{15} M_{\odot}/h$.  Eq.~(\ref{eq.Npoint}) only includes the 1-halo contribution to the $N^{\mathrm{th}}$ moment at zero lag --- as noted, we have neglected contributions due to clustering.  Using the bias model of~\citep{Tinkeretal2010}, we find that including the 2-halo term typically increases $\langle T^2 \rangle$ by only $1$--$2$\%.  Moreover, the 2- and 3-halo contributions to $\langle T^3 \rangle$ should be even less significant, since the higher moments are dominated by regions increasingly near the center of each cluster (where $|T|$ is larger).  Thus, we include only the 1-halo term in all calculations, as given in Eq.~(\ref{eq.Npoint}).

Note that we define $M$ to be the virial mass, that is, the mass enclosed within a radius $r_{vir}$~\cite{Bryan-Norman1998}:
\be
\label{eq.rvir}
r_{vir} = \left( \frac{3 M}{4 \pi \Delta_{cr}(z) \rho_{cr}(z)} \right)^{1/3} \,,
\ee
where $\Delta_{cr}(z) = 18 \pi^2 + 82(\Omega(z)-1) - 39(\Omega(z)-1)^2$, $\Omega(z) = \Omega_m(1+z)^3/(\Omega_m(1+z)^3+\Omega_{\Lambda})$, and $\rho_{cr}(z) = 3H^2(z)/8\pi G$ is the critical density at redshift $z$.  However, some of the pressure profiles that we use below are specified as a function of the spherical overdensity mass rather than the virial mass.  Thus, when necessary, we use the NFW density profile~\cite{NFW1997} and the concentration-mass relation from~\citep{Duffyetal2008} to convert the virial mass to a spherical overdensity mass $M_{\delta,c/d}$, where $M_{\delta,c}$ ($M_{\delta,d}$) is the mass enclosed within a sphere of radius $r_{\delta,c}$ ($r_{\delta,d}$) such that the enclosed density is $\delta$ times the critical (mean matter) density at redshift $z$.  To be completely explicit, this procedure involves solving the following non-linear equation for $r_{\delta,c}$ (or $r_{\delta,d}$):
\be
\label{eq.rvir2}
\int_0^{r_{\delta,c}} 4 \pi r'^2 \rho_{\mathrm{NFW}} (r', M_{vir}, c_{vir}) dr' = \frac{4}{3} \pi r_{\delta,c}^3 \rho_{cr}(z) \delta
\ee
where $c_{vir} \equiv r_{vir}/r_s$ is the concentration parameter ($r_s$ is the NFW scale radius) and one replaces the critical density $\rho_{cr}(z)$ with the mean matter density $\rho_m(z)$ in this equation in order to obtain $r_{\delta,d}$ instead of $r_{\delta,c}$.  After solving Eq.~(\ref{eq.rvir2}) to find $r_{\delta,c}$, $M_{\delta,c}$ is simply calculated via $M_{\delta,c} = \frac{4}{3} \pi r_{\delta,c}^3 \rho_{cr}(z) \delta$.

From Eqs.~(\ref{eq.tSZdef}) and (\ref{eq.Npoint}), it is clear that two ingredients are required for the tSZ moment
calculation, given a specified cosmology:  (1) the halo mass function $dn(M,z)/dM$ and (2) the
electron pressure profile $P_e(r,M,z)$ for halos of mass $M$ at
redshift $z$.  The calculation thus naturally divides into a cosmology-dependent component (the mass function) and an ICM-dependent component (the pressure profile).  Because the small-scale baryonic physics responsible for the details of the ICM pressure profile is decoupled from the large-scale physics described by the cosmological parameters ($\sigma_8, \Omega_m$, ...), it is conventional to determine the pressure profile from numerical cosmological hydrodynamics simulations (or observations of galaxy clusters), which are run for a fixed cosmology.  One can then take this pressure profile and compute its predictions for a different cosmology by using the halo mass function appropriate for that cosmology.  We follow this approach below.

For all the calculations in this paper, we use the $M_{200,d}$ halo
mass function of~\citep{Tinkeretal2008} with the redshift-dependent
parameters given in their Eqs.~(5)--(8); we will hereafter refer to
this as the Tinker mass function.  However, the tSZ moments are somewhat sensitive
to the particular choice of mass function used in the calculation.  As a test, we perform identical
calculations using the $M_{400,d}$ halo mass function of~\citep{Tinkeretal2008}.  Using our fiducial
cosmology and the Battaglia pressure profile (see below), we find that the $M_{400,d}$ mass function
gives a result for $\langle T^2 \rangle$ that is $11$\% higher than that found using the $M_{200,d}$ mass function,
while the result for $\langle T^3 \rangle$ is $22$\% higher.  In general, higher tSZ moments are
more sensitive to changes in the mass function, as they are dominated by progressively higher-mass halos
that live in the exponential tail of the mass function.  While these changes are non-negligible for upcoming high precision
measurements, they are smaller than those caused by changes in the choice of pressure profile; hence, for the remainder of this paper, we use the $M_{200,d}$ Tinker mass function for all calculations.

Our calculations include four different pressure profiles from~\citep{Battagliaetal2011b}, \citep{Arnaudetal2010}, and \citep{Komatsu-Seljak2002} (and additional results using the profile from~\citep{Bhattacharyaetal2012}), which we briefly describe in the following.  We consider two profiles derived from the simulations of~\citep{Battagliaetal2010}, which are reported in~\citep{Battagliaetal2011b}.  The first, which we hereafter refer to as the Battaglia profile, is derived from hydrodynamical simulations that include radiative cooling, star formation, supernova feedback, and feedback from active galactic nuclei (AGN).  These feedback processes tend to lower the gas fraction in low-mass clusters, as gas is blown out by the injection of energy into the ICM.  The smoothed particle hydrodynamics used in these simulations also captures the effects of non-thermal pressure support due to bulk motions and turbulence.  The second profile that we use from~\citep{Battagliaetal2011b}, which we hereafter refer to as the Battaglia Adiabatic profile\footnote{N.\ Battaglia, priv.\ comm. (to be included in~\citep{Battagliaetal2011b})}, is derived from hydrodynamical simulations with all forms of feedback turned off.  This setup leads to a non-radiative adiabatic cluster model with only formation shock heating present.  Thus, in these simulations, the gas fraction in low-mass clusters is close to the cosmological value, $\Omega_b/\Omega_m$ (as is the case in high-mass clusters).  This leads to much higher predicted amplitudes for the tSZ moments (see below).  Note that the Battaglia profile is specified as a function of $M_{200,c}$, while the Battaglia Adiabatic profile is given as a function of $M_{500,c}$. %Note that both the Battaglia and Battaglia Adiabatic profiles are determined from cosmological hydrodynamics simulations run with $\sigma_8 = 0.8$.  However, as mentioned above, the small-scale physics responsible for the pressure profile details is decoupled from the large-scale cosmological background.  Thus, we compute the predictions of this pressure profile for other cosmologies by simply re-computing the mass function appropriately (as done in~\citep{Battagliaetal2011b}).

In addition, we use the ``universal'' pressure profile derived in~\citep{Arnaudetal2010}, which we hereafter refer to as the Arnaud profile.  This profile is obtained from a combination of \emph{XMM-Newton} observations of massive, $z<0.2$ clusters and hydrodynamical simulations that include radiative cooling and some feedback processes (though not AGN feedback).  The simulations are used to extend the profile beyond $R_{500}$, due to the lack of X-ray photons at large radii.  Importantly, the normalization of this profile is obtained using hydrostatic equilibrium (HSE)-based estimates of the mass $M_{500,c}$, which are known to be biased low by $\approx 10-15$\%~\cite{Piffaretti-Valdarnini2008}.  Thus, following~\cite{Shawetal2010}, we use a HSE-bias correction of $13$\% in our calculations with this profile, that is, we set $M_{500,c}^{HSE} = 0.87 M_{500,c}$.  This correction slightly lowers the amplitude of the tSZ moments for this profile.

We also use the profile derived in~\citep{Komatsu-Seljak2002}, which we hereafter refer to as the Komatsu-Seljak (or K-S) profile.  This profile is derived analytically under some simplifying assumptions, including HSE, gas tracing dark matter in the outer regions of clusters, and a constant polytropic equation of state for the ICM gas.  In particular, it includes no feedback prescriptions or sources of non-thermal pressure support.  The gas fraction in low-mass clusters (indeed, in all clusters) is thus equal to the cosmological value, leading to high predicted amplitudes for the tSZ moments. This profile is now known to over-predict the tSZ power spectrum amplitude~\citep{Dunkleyetal2011,Reichardtetal2011}, but we include it here as an extreme example of the possible ICM physics scenarios.  It is specified as a function of the virial mass $M$ as defined in Eq.~(\ref{eq.rvir}).

Finally, we present results computed using the fiducial profile from~\citep{Bhattacharyaetal2012}, which have been graciously provided by the authors of that study.  This profile is based on that of~\cite{Shawetal2010}, and thus we hereafter refer to this as the Shaw profile.  It is derived from an analytic ICM model that accounts for star formation, feedback from supernovae and AGN, and non-thermal pressure support.  Its tSZ predictions are generally similar to those of the Battaglia profile.  In particular, it also predicts a suppression of the gas fraction in low-mass clusters due to feedback processes.

\begin{table}
\begin{tabular}{c | cc | cc}
\label{tab.scalings}
Profile & $A_2$ [$\mu$K$^2$] & $A_3$ [$\mu$K$^3$] & $\alpha_2$ & $\alpha_3$ \\
\hline
Arnaud & $20.5$ & $-1790$ & $7.9$ & $11.5$ \\
Battaglia & $22.6$ & $-1660$ & $7.7$ & $11.2$ \\
Batt.\ Adiabatic & $47.1$ & $-3120$ & $6.6$ & $9.7$ \\
Komatsu-Seljak & $53.0$ & $-3040$ & $7.5$ & $10.6$ \\
Shaw & $23.8$ & $-1610$ & $7.9$ & $10.7$ \\
\end{tabular}
\caption{Amplitudes and power-law scalings with $\sigma_8$ for the tSZ variance and skewness, as defined in Eq.~(\ref{eq.powerlaw}). The first column lists the pressure profile used in the calculation (note that all calculations use the Tinker mass function).  The amplitudes are specified at $\sigma_8 = 0.817$, the WMAP5 maximum-likelihood value.  All results are computed at $\nu = 150$ GHz.}
\end{table}

Overall, then, we have two profiles that include a variety of detailed feedback prescriptions (Battaglia and Shaw), one profile based primarily on an empirical fit to X-ray data (Arnaud), and two profiles for which the ICM is essentially in HSE (Battaglia Adiabatic and Komatsu-Seljak).  The Arnaud, Battaglia, and Shaw models all agree reasonably well with X-ray observations of massive, low-redshift clusters~\cite{Sunetal2011}.  The Komatsu-Seljak profile is somewhat discrepant for these high-mass systems, but disagrees more significantly with the predictions of the feedback-based profiles for low-mass groups and clusters.  The predicted gas fraction in these low-mass objects is a major source (along with non-thermal pressure support) of the difference in the tSZ predictions between the feedback- or X-ray-based models (Arnaud, Battaglia, and Shaw) and the adiabatic models (Battaglia Adiabatic and Komatsu-Seljak).

\begin{figure}
\label{fig.twopt}
  \begin{center}
    \includegraphics[width=\columnwidth]{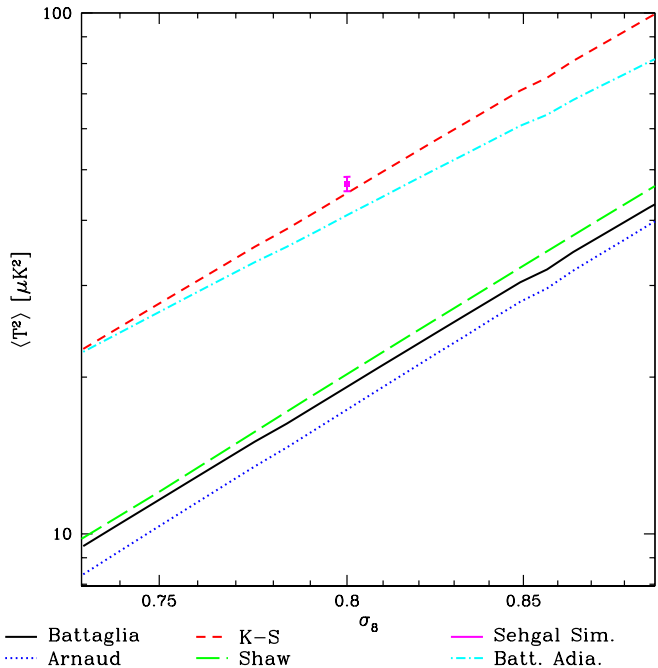}
    \caption{The tSZ variance versus $\sigma_8$ obtained from five different pressure profiles (using the Tinker mass function) and one direct simulation measurement. The scalings with $\sigma_8$ are similar for all the models: $\langle T^2 \rangle \propto \sigma_8^{6.6-7.9}$.  It is evident that $\langle T^2 \rangle$ is very sensitive to $\sigma_8$, but the scatter due to uncertainties in the ICM astrophysics (which is greater than twice the signal for large $\sigma_8$) makes precise constraints from this quantity difficult.}
  \end{center}
\end{figure}

We also analyze a tSZ simulation from \cite{Sehgaletal2010} that covers an octant of the sky, providing a nontrivial test of our calculations.  We hereafter refer to this as the Sehgal simulation.  The simulated data was produced by populating a large dark matter $N$-body simulation with gas according to a polytropic cluster model that also includes some feedback prescriptions.  However, the model requires all clusters --- including low-mass objects --- to contain enough gas to reach the cosmological value of the gas fraction.  Note that the simulation results include tSZ signal from components that are not accounted for in the halo model-based calculations, including substructure within halos, deviations from the globally averaged pressure profile, and diffuse emission from the intergalactic medium.  These effects are expected to contribute to the tSZ power spectrum at the  $\approx 10-20$\% level at high-$\ell$~\citep{Battagliaetal2011b}. The simulation was run using $\sigma_8 = 0.8$, with the other $\Lambda$CDM parameters taking values consistent with WMAP5.  Since an average pressure profile has not been derived from this simulation, we cannot re-compute its predictions for different cosmologies, and thus it is presented as a single data point in the figures throughout this paper.

\begin{figure}
\label{fig.threept}
  \begin{center}
    \includegraphics[width=\columnwidth]{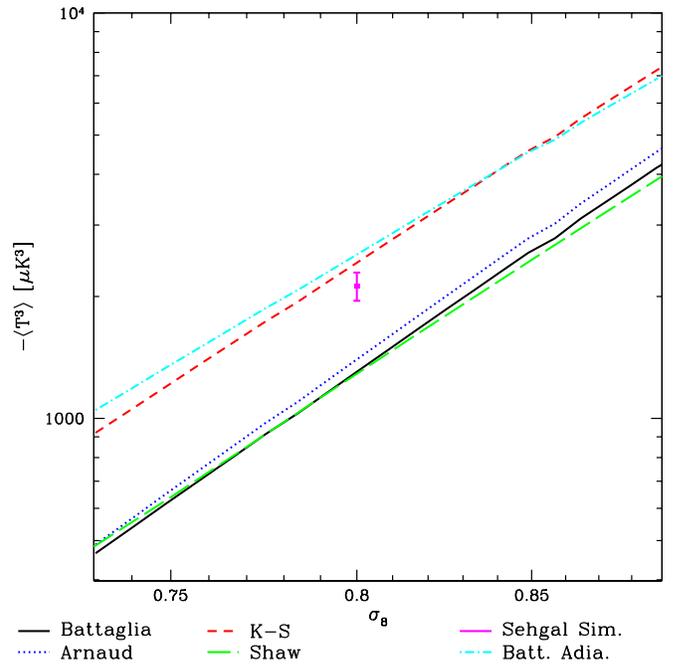}
    \caption{Similar to Fig.~1, but now showing the tSZ skewness.  The scalings with $\sigma_8$ are again similar for all the models: $\langle T^3 \rangle \propto \sigma_8^{9.7-11.5}$. As for the variance, the sensitivity to $\sigma_8$ is pronounced, but degraded due to uncertainties in the ICM astrophysics (as represented by the different choices of pressure profile).}
  \end{center}
\end{figure}

\subsection{Results}
We compute Eq.~(\ref{eq.Npoint}) for the variance ($N=2$) and the unnormalized skewness ($N=3$) for each of the profiles while varying $\sigma_8$ and keeping the other
cosmological parameters fixed.  The results are shown in Figs.~1 and 2.  (See Appendix B for results involving the unnormalized kurtosis, i.e., $N=4$.)  We find that the scalings of the variance and skewness with $\sigma_8$ are well-described by power-laws for each of these profiles,
\be
\label{eq.powerlaw}
\langle T^{2,3} \rangle = A_{2,3} \left( \frac{\sigma_8}{0.817} \right)^{\alpha_{2,3}} \,,
\ee
where we have normalized to the WMAP5 value of $\sigma_8$.  The amplitudes $A_{2,3}$ and scalings $\alpha_{2,3}$ for each of the pressure profiles are given in Table I. The scalings are similar for all the pressure profiles. Note that the slightly steeper scalings for the  Battaglia and Shaw profiles are due to their inclusion of feedback processes that suppress the tSZ signal from low-mass clusters (the Arnaud profile also suppresses the signal from these objects, though it is primarily based on a fit to higher-mass clusters).  Thus, $\langle T^2 \rangle$ and $\langle T^3 \rangle$ for these profiles are dominated by higher-mass (rarer) objects than for the Komatsu-Seljak or Battaglia Adiabatic profiles, and they are correspondingly more sensitive to $\sigma_8$.  Despite the steep scalings, it is clear in Figs.~1 and 2 that the systematic uncertainty due to the unknown ICM astrophysics significantly degrades any potential constraints that could be derived from these observables.  We address possible ways around this problem in the next section.

\begin{figure}
\label{fig.twoptvsMmax}
  \begin{center}
    \includegraphics[width=\columnwidth]{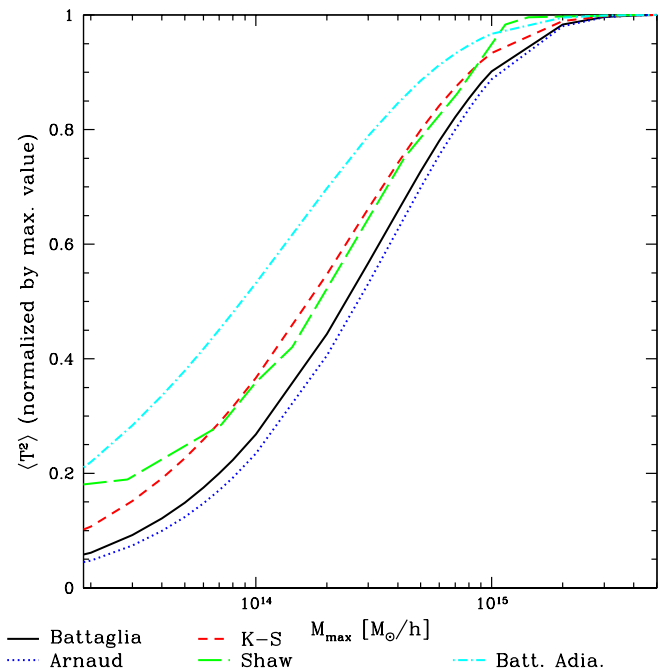}
    \caption{Fraction of the tSZ variance contributed by clusters with virial mass $M < M_{\mathrm{max}}$.}
  \end{center}
\end{figure}

We also investigate the characteristic mass and redshift ranges responsible
for the variance and skewness signals.  Figs.~3 and 4 show the fraction of the
variance and skewness signals contributed by clusters with virial mass $M < M_{\mathrm{max}}$.
We find that the variance receives $\approx 50-60$\% of its amplitude from clusters with
$M < 2$--$3 \times 10^{14} \,\, \mathrm{M}_{\odot}/h$, while the
skewness receives only $\approx 20-40$\% of its amplitude
from these less massive objects.  These results vary for the
different profiles, as the different feedback prescriptions suppress
the amplitude contributed by low-mass clusters by different amounts.
More massive clusters are dominated by gravitational heating
and are thus less sensitive to energy input via feedback from AGN, turbulence, and other
mechanisms \cite{Battagliaetal2011a,Shawetal2010}.  Since the
Komatsu-Seljak and Battaglia Adiabatic profiles include no feedback, the amplitudes of their
variance and skewness are somewhat more weighted toward low-mass objects.  In all cases, though,
these results indicate that the ICM astrophysics underlying the tSZ
skewness is much better constrained by observation than that responsible for
much of the variance (or power spectrum), as the skewness amplitude is dominated by significantly more massive objects which have been better studied observationally.  This explains the smaller scatter seen between the different profiles in Fig.~2 as compared to Fig.~1: the skewness signal is dominated by objects for which the ICM astrophysics is less uncertain.

%In addition, we find that the variance receives $\approx
%25$\% of its amplitude from groups and clusters at $z>1$, while the
%skewness receives only $\approx 15$\% of its amplitude
%from these high-redshift objects.  Although this result is not as
%dramatic as the mass characterization above, it does imply that a
%greater fraction of the variance than the skewness originates from clusters that have been studied using X-rays, lensing, and other techniques.  On the other hand, most of the objects comprising the tSZ variance (or power spectrum) have not been directly observed.
In addition, we find that the variance receives $\approx
25$\% (Arnaud/Battaglia) -- $45$\% (Battaglia Adiabatic) of its amplitude from groups and clusters at $z>1$, while the skewness receives only $\approx 15$\% (Arnaud/Battaglia) -- $35$\% (Battaglia Adiabatic) of its amplitude
from these high-redshift objects.  (The Komatsu-Seljak results lie in the middle of these ranges.)  These results imply that a greater fraction of the tSZ skewness than the variance originates from clusters that have been studied using X-rays, lensing, and other techniques.  On the other hand, many of the objects comprising the variance (or power spectrum) have not been directly observed.

\begin{figure}
\label{fig.threeptvsMmax}
  \begin{center}
    \includegraphics[width=\columnwidth]{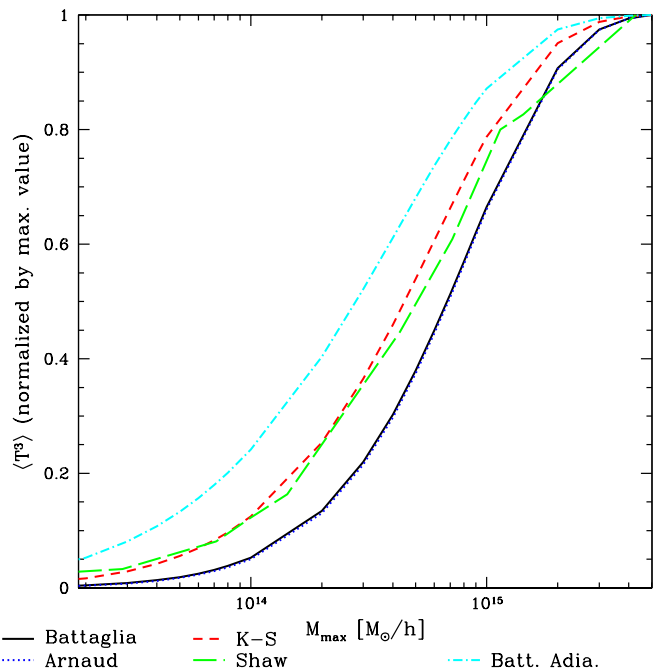}
    \caption{Fraction of the tSZ skewness contributed by clusters with virial mass $M < M_{\mathrm{max}}$.  Comparison with Fig.~3 indicates that the skewness signal arises from higher-mass clusters than those that comprise the variance.}
  \end{center}
\end{figure}

As it depends sensitively on $\sigma_8$ and is less affected by uncertainties in astrophysical modeling than the power spectrum, the tSZ skewness is a powerful cosmological probe. Nevertheless, despite the signal originating from characteristically higher-mass, lower-redshift clusters, the amplitude of the skewness is still quite uncertain due to the poorly understood astrophysics of the ICM, as can be seen in Fig.~2.  We investigate methods to solve this problem in the remainder of this paper. %In subsequent sections, we discuss methods to minimize the errors due to ICM astrophysics uncertainty on cosmological information extracted from the tSZ signal.

\begin{figure}
\label{fig.sig8ratiocancel}
  \begin{center}
    \includegraphics[width=\columnwidth]{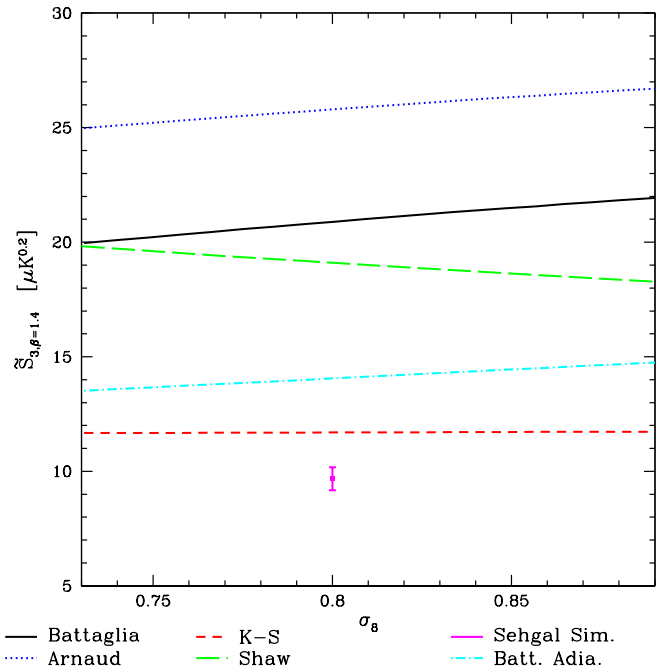}
    \caption{The rescaled skewness $\tilde{S}_{3,\beta}$ for $\beta=1.4$ plotted against $\sigma_8$.  It is evident that this statistic is nearly independent of $\sigma_8$, as expected based on the scalings in Table~I.  However, it is still dependent on the ICM astrophysics, as represented by the different pressure profiles.  Thus, this statistic can be used for determining the correct gas physics model.}
  \end{center}
\end{figure}

\section{Isolating the Dependence on ICM Astrophysics}
\label{sec:isolate}
%\jch{remember to mention how we got the Shaw results! also, keep the $b_{\pi}$ stuff and use it to emphasize that if naive
%  pressure biases were equal, the 1.4-1.5 power would cancel both the
% sigma8 dependence AND the ICM dependence -- it does not, because the
%  two statistics are probing different mass scales / different biases; also, emphasize our lack of understanding of the 0.7 ratio trick statistic, and %how it is likely to be affected by changes in mass function, non-thermal support (if implemented in the Shaw manner), etc.}
As we have described, the cosmological utility of tSZ measurements is limited by their sensitivity to unknown, non-linear astrophysical processes in the ICM, despite their high sensitivity to the underlying cosmology. However, this theoretical uncertainty can be minimized by combining measurements of the tSZ skewness with measurements of the tSZ power spectrum or variance. The use of multiple probes provides an additional statistical handle on both ICM astrophysics and cosmology.

\begin{figure}
\label{fig.sig8ratiocancelvsMmax}
  \begin{center}
    \includegraphics[width=\columnwidth]{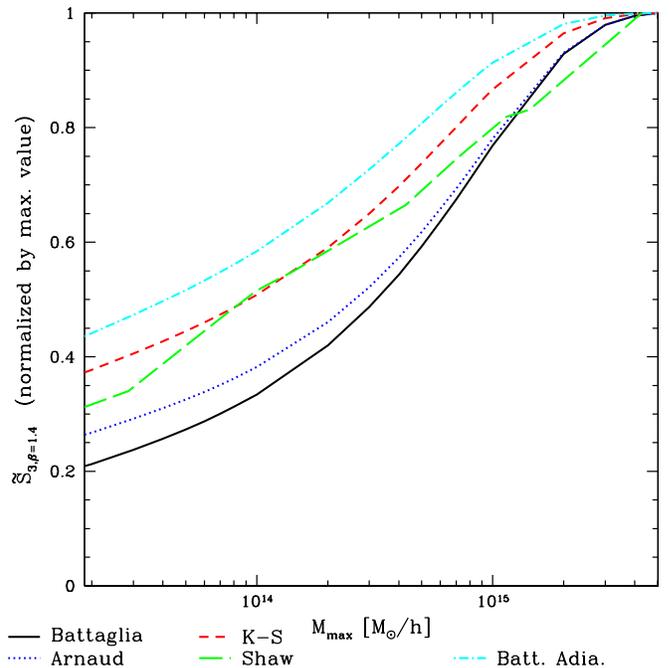}
    \caption{Fraction of the rescaled skewness $\tilde{S}_{3,\beta}$ for $\beta=1.4$ contributed by clusters with virial mass $M < M_{\mathrm{max}}$.  A significant fraction of the signal comes from low-mass objects --- even more so than the variance (Fig.~3).  One can thus interpret this statistic as a measure of the gas fraction in low-mass clusters: pressure profiles that yield $f_{gas} \approx \Omega_b/\Omega_m$ in low-mass objects lead to a small value for this statistic (since they give much larger values of $\langle T^2 \rangle$), while pressure profiles that include significant feedback effects --- thus lowering the gas fraction in low-mass objects --- lead to larger values for this statistic.}
  \end{center}
\end{figure}

If the tSZ variance and skewness depend differently on the ICM astrophysics and background cosmology, it may be possible to construct a statistic that ``cancels'' the dependence on one or the other.  Here, we focus on a statistic that cancels the dependence on cosmology, but preserves a dependence on the ICM astrophysics.  For an alternative approach that attempts to directly cancel the ICM astrophysics dependence, see Appendix A.  In particular, we consider the following statistic, which we call the ``rescaled skewness'':
\be
\label{eq.ratiostat}
\tilde{S}_{3,\beta} = \frac{|\langle T^3 \rangle|}{\langle T^2 \rangle^{\beta}} \,,
\ee
where the value of $\beta$ is left to be determined.  %As a first test, we compute Eq.~(\ref{eq.ratiostat}) for a value of $\beta$ that should cancel the dependence on $\sigma_8$. 
From the results in Table~I, we know that $\langle T^2 \rangle \propto \sigma_8^{6.6-7.9}$ and $\langle T^3 \rangle \propto \sigma_8^{9.7-11.5}$ for these pressure profiles.  Thus, if we choose $\beta = 1.4$, then the resulting statistic will cancel the dependence on $\sigma_8$, and thus should be nearly independent of the background cosmology, since $\sigma_8$ is the dominant cosmological parameter for the tSZ observables.  Nonetheless, some dependence on the ICM astrophysics may remain.  We investigate this dependence in Fig.~5.  It is evident that the rescaled skewness with $\beta=1.4$ is nearly independent of $\sigma_8$, as expected based on the scalings.  However, this statistic still shows a dependence on the gas physics model --- for $\sigma_8=0.817$, the scatter in $\tilde{S}_{3,\beta=1.4}$ between the different pressure profiles is $\approx 35$\%.  For the same value of $\sigma_8$, the scatter in $\langle T^2 \rangle$ is $\approx 50$\% between these profiles, while for $\langle T^3 \rangle$ it is $\approx 35$\%.  Thus, the scatter in this statistic is essentially identical to that in the skewness. The rescaled skewness with $\beta=1.4$ hence has a very useful property: because it only depends weakly on $\sigma_8$, it can be used to determine the correct ICM gas physics model, nearly independent of the background cosmology.  Although the value of $\beta$ could be chosen slightly differently to minimize the small residual dependence on $\sigma_8$ for any of the individual pressure profiles, we find that $\beta=1.4$ is the best model-independent choice available, especially given current levels of observational precision and theoretical uncertainty.  After fixing the value of $\beta$, one can proceed to measure this statistic from the data (see $\S$IV).

Moreover, the ordering of the results in Fig.~5 suggests a possible interpretation of this statistic.  The pressure profiles for which $f_{gas} \approx \Omega_b/\Omega_m$ in all halos --- including low-mass groups and clusters --- yield low values for $\tilde{S}_{3,\beta=1.4}$, while the profiles with feedback prescriptions that lower $f_{gas}$ in low-mass objects yield high values of this statistic.  This can be explained by the fact that the additional contributions from low-mass clusters lead to higher values of $\langle T^2 \rangle$ in the former set of profiles, while $\langle T^3 \rangle$ is not significantly affected, since it is dominated by contributions from more massive objects.  Looking at Eq.~(\ref{eq.ratiostat}), it then follows that the profiles without $f_{gas}$ suppression have lower values for $\tilde{S}_{3,\beta=1.4}$.  Thus, the rescaled skewness with $\beta=1.4$ is a measure of the typical gas fraction in low-mass groups and clusters.

We verify this interpretation by calculating the characteristic mass range responsible for the $\tilde{S}_{3,\beta=1.4}$ signal.  Fig.~6 shows the fraction of this signal contributed by clusters with virial mass $M < M_{\mathrm{max}}$.  As anticipated, it receives significant contributions from very low-mass objects: $\approx 30-50$\% of the amplitude comes from clusters with $M < 6$--$7 \times 10^{13} \,\, \mathrm{M}_{\odot}/h$ (this depends quite strongly on the particular profile used, as seen in the figure).  A measurement of this quantity will thus constrain the average amount of gas in these low-mass, high-redshift objects, which have not yet been observed by other techniques.  We present a first application of this technique to data from ACT and SPT in the following section.
%%Blake: I think it is definitely true that S1.4 is sensitive to differences in fgas. what i do not think is clear is that it is not sensitive to other things as well, though it is likely.

\section{Application to ACT and SPT Data}
\label{sec:data}
ACT and SPT have measured the tSZ signal in maps of the microwave sky with arcminute angular resolution.  In order to apply the rescaled skewness with $\beta=1.4$ to the data, we rephrase this statistic to match the observational quantities as closely as possible.  In particular, we use the SPT measurement of $C_{3000}$~\cite{Reichardtetal2011}, the amplitude of the tSZ power spectrum at $\ell=3000$, and the ACT measurement of $\langle \tilde T^3 \rangle$~\cite{Wilsonetal2012}, the filtered tSZ skewness (defined more precisely below).

We use $C_{3000}$ rather than $\langle T^2 \rangle$ in order to circumvent uncertainties that arise when converting between the tSZ power spectrum and variance.  ACT and SPT effectively measure the amplitude of the tSZ power spectrum only at scales of order $\ell=3000$, and therefore a template for $C_{\ell}$ must be used in order to calculate the tSZ variance from this measurement.  Such a template relies on an ICM physics model, and since this model is what we are hoping to constrain, we circumvent the calculation of the variance by working directly with $C_{3000}$ instead of $\langle T^2 \rangle$.  We use the SPT measurement of $C_{3000}$ due to its higher signal-to-noise ratio (SNR) than the measurement from ACT.  Finally, note that SPT measures this quantity at an effective frequency of $152.9$ GHz~\cite{Reichardtetal2011}, and thus we perform the relevant calculations at this frequency.

\begin{figure}
\label{fig.dataplot}
  \begin{center}
    \includegraphics[width=\columnwidth]{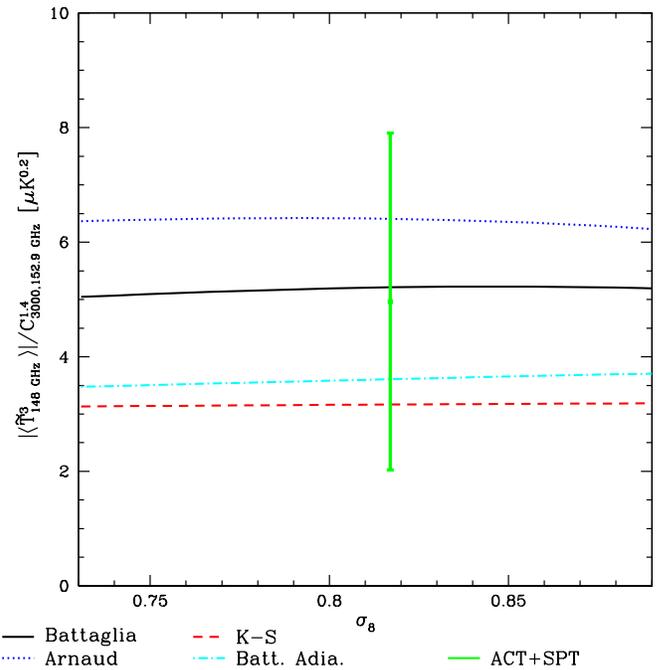}
    \caption{Similar to Fig.~5, but calculated in terms of observed tSZ statistics, namely, $C_{3000}$ (the amplitude of the tSZ power spectrum at $\ell = 3000$) and $\langle \tilde T^3 \rangle$ (the filtered skewness as defined in \cite{Wilsonetal2012}).  The green point shows the current constraint using the SPT measurement of $C_{3000}$ \cite{Reichardtetal2011} and the ACT measurement of $\langle \tilde T^3 \rangle$ \cite{Wilsonetal2012}.}
  \end{center}
\end{figure}

ACT recently reported the first detection of a higher-point tSZ observable: the filtered skewness, $\langle \tilde T^3 \rangle$~\cite{Wilsonetal2012}.  This quantity is similar to the tSZ skewness that we calculate above, but its value has been modified by an $\ell$-space filter applied to the ACT maps, as well as a temperature fluctuation cut-off used to remove outlying pixels in the ACT maps.  We explicitly account for these steps in our calculations.  First, we Fourier transform the projected temperature decrement profile in Eq.~(\ref{eq.tSZdef}) and apply the $\ell$-space filter used in~\cite{Wilsonetal2012} to each ``cluster'' of mass $M$ and redshift $z$ in the integrals of Eq.~(\ref{eq.Npoint}).  We then inverse Fourier transform to obtain the filtered temperature decrement profile for each cluster in real space.  Second, we place each cluster in an idealized ACT pixel and compute the observed temperature decrement, accounting carefully for geometric effects that can arise depending on the alignment of the cluster and pixel centers.  If the observed temperature decrement exceeds the $12\sigma$ cut-off used in~\cite{Wilsonetal2012}, then we discard this cluster from the integrals.  We thus replicate the data analysis procedure used in the ACT measurement as closely as possible.  The net effect is to reduce the value of the tSZ skewness by up to $90-95$\%.  The reduction comes almost entirely from the $\ell$-space filtering; the cut-off used in the second step only has a small effect.  Finally, note that the ACT measurement is at an effective frequency of $148$ GHz~\cite{Dunkleyetal2011}, and thus we perform the relevant calculations at this frequency.

\begin{figure}
\label{fig.2ptvsMnu}
  \begin{center}
    \includegraphics[width=\columnwidth]{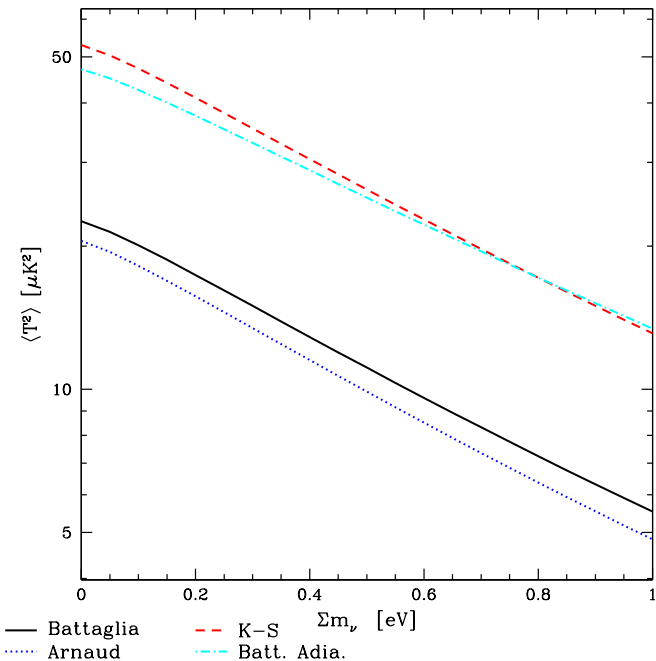}
    \caption{The tSZ variance versus the sum of the neutrino masses $\Sigma m_{\nu}$, with $\Delta_{\mathcal{R}}^2 = 2.46 \times 10^{-9}$ (its
WMAP5 value).  The dependence is not precisely captured by a simple scaling as for $\sigma_8$ (note that the axes are log-linear), but the curves are well-fit by quadratic polynomials.}
  \end{center}
\end{figure}

We use these calculations to construct a statistic analogous to $\tilde{S}_{3,\beta=1.4}$, but defined in terms of the observational quantities: $|\langle \tilde T^3 \rangle|/C_{3000}^{1.4}$.  In order for this statistic to work in the manner seen in Fig.~5, we thus require the scalings of $C_{3000}$ and $\langle \tilde T^3 \rangle$ with $\sigma_8$ to be close to those reported in Table~I for $\langle T^2 \rangle$ and $\langle T^3 \rangle$, respectively.  Our calculations verify this claim, as can be seen immediately in Fig.~7.

Fig.~7 shows the result from ACT and SPT plotted against the results of our theoretical calculations.  Unfortunately, the error bar on the data point is too large to deduce a preference for any particular pressure profile (though the central value is closer to the profiles with significant feedback than those with an unsuppressed value of $f_{gas}$ in all halos).  Note that the error bar includes significant contributions due to cosmic variance (i.e., sample variance) resulting from the limited sky coverage of ACT and SPT, as detailed in~\cite{Wilsonetal2012} and \cite{Reichardtetal2011}.  The nearly full-sky results from \emph{Planck} will have far smaller cosmic variance contributions to the error.  Moreover, the error is currently dominated by the uncertainty on $\langle \tilde T^3 \rangle$, which will greatly decrease in the near future --- an upcoming SPT measurement should increase the SNR on $\langle \tilde T^3 \rangle$ by a factor of 3~\cite{Bhattacharyaetal2012}.

Nonetheless, other theoretical issues must still be overcome in order to fully characterize this technique.  In particular, the tSZ power spectrum receives non-negligible contributions (up to $15$\% at $\ell=3000$) due to deviations about the mean pressure profile found in simulations~\cite{Battagliaetal2011b}.  The tSZ skewness likely receives similar contributions, which are not accounted for in our halo model-based calculations.  Future studies incorporating the results of hydrodynamical cosmological simulations will be necessary to fully understand this method.  Fig.~7 is simply a proof of concept for this technique, which may soon give interesting constraints on the ICM astrophysics of low-mass groups and clusters.  By using the constrained pressure profile to interpret measurements of the power spectrum or skewness, it will be possible to derive stronger constraints on $\sigma_8$ from tSZ measurements.

\begin{figure}
\label{fig.3ptvsMnu}
  \begin{center}
    \includegraphics[width=\columnwidth]{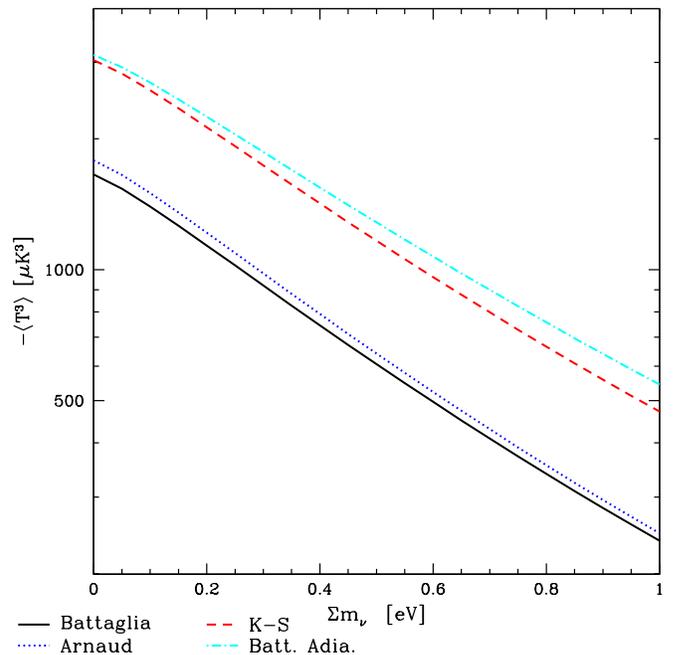}
    \caption{The tSZ skewness versus the sum of the neutrino masses $\Sigma m_{\nu}$, with $\Delta_{\mathcal{R}}^2 = 2.46 \times 10^{-9}$ (its
WMAP5 value).  The dependence is not precisely captured by a simple scaling as for $\sigma_8$ (note that the axes are log-linear), but the curves are well-fit by cubic polynomials.}
  \end{center}
\end{figure}

\section{Future Cosmological Constraints}
\label{sec:constraints}
\subsection{Extension to other parameters}
While we have focused on $\sigma_8$, the techniques that we have outlined are in principle sensitive to many cosmological parameters.  These include the standard parameters $\Omega_b h^2$, to which the tSZ signal is somewhat sensitive (though this parameter can be very well constrained by the primordial CMB), and $\Omega_m$, to which the tSZ signal is not particularly sensitive (especially in the range $0.15<\Omega_m<0.4$)~\cite{Komatsu-Seljak2002}.  Here, we instead focus on currently unknown parameters, to which the tSZ moments are sensitive through the halo mass function.  Such parameters include the sum of the neutrino masses $\Sigma m_{\nu}$, the non-Gaussianity parameters (e.g., $f_{NL}$), and the dark energy equation of state $w(z)$.  All of these parameters modify the number of massive clusters in the low-redshift universe: for example, non-zero neutrino masses suppress the number of clusters, while positive $f_{NL}$ provides an enhancement.  Thus, these parameters also suppress or enhance the amplitude of the tSZ moments.

\begin{figure}
\label{fig.2ptvsfNL}
  \begin{center}
    \includegraphics[width=\columnwidth]{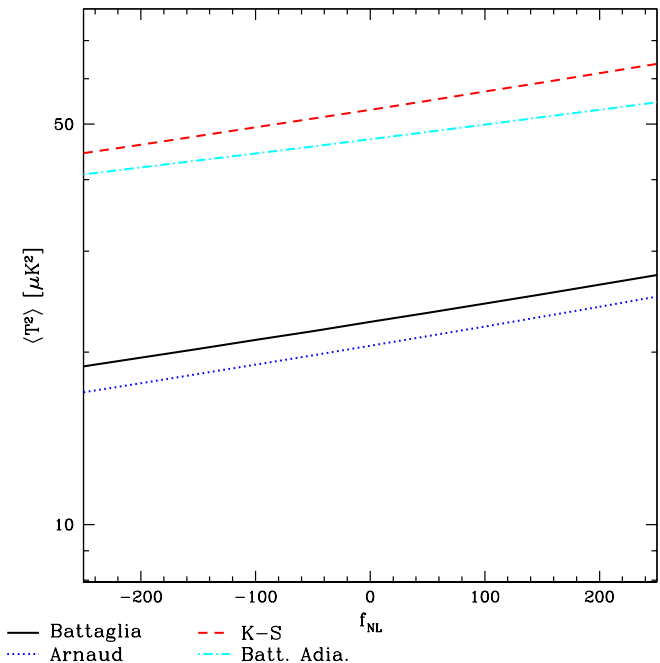}
    \caption{The tSZ variance versus $f_{NL}$, with $\Delta_{\mathcal{R}}^2 = 2.46 \times 10^{-9}$ (its
WMAP5 value).  The dependence is not precisely captured by a simple scaling as for $\sigma_8$ (note that the axes are log-linear), but the curves are well-fit by quadratic polynomials.}
  \end{center}
\end{figure}

As a first example, we compute the tSZ variance and skewness for a fixed WMAP5 background cosmology with massive neutrinos added.  To be clear, we fix $\Delta_{\mathcal{R}}^2 = 2.46 \times 10^{-9}$ (its WMAP5 value), not $\sigma_8 = 0.817$, as the presence of massive neutrinos will
lead to a lower value of $\sigma_8$ inferred at the low redshifts from which the tSZ signal originates.  We compute the change in the mass function due to the neutrinos following a prescription similar to that of \cite{Ichiki-Takada2011}, except that we input the suppressed linear theory matter power spectrum to the Tinker mass function rather than that of \cite{Bhattacharyaetal2011}, as was done in \cite{Ichiki-Takada2011}.  A more precise recipe can be found in \cite{Brandbygeetal2010,Marullietal2011}, but this approach should capture the relevant physical effects.

The results of this analysis are shown in Figs.~8 and 9.  As expected, the tSZ variance and skewness both exhibit sensitivity to $\Sigma m_{\nu}$, although it is currently overwhelmed by the uncertainty in the ICM astrophysics. However, using the method outlined in the previous section to constrain the ICM astrophysics and thus break its degeneracy with cosmology, a measurement of the sum of the neutrino masses with the tSZ effect might be feasible.

\begin{figure}
\label{fig.3ptvsfNL}
  \begin{center}
    \includegraphics[width=\columnwidth]{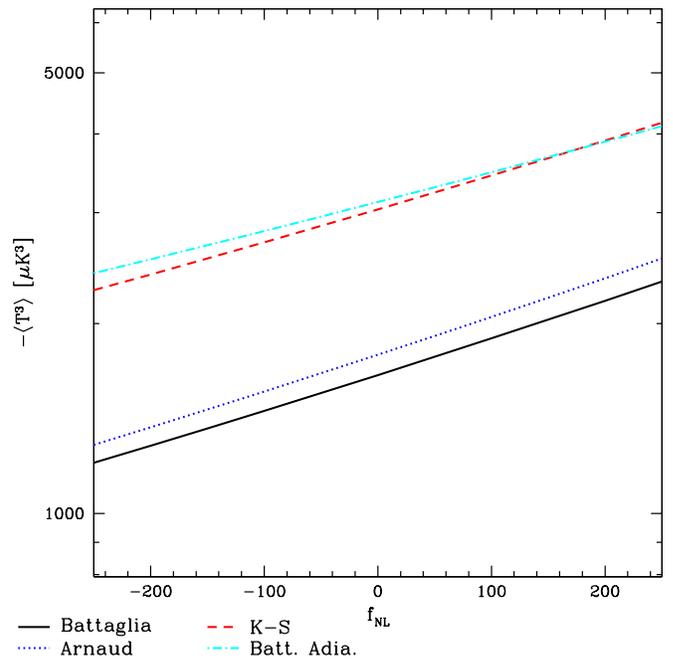}
    \caption{The tSZ skewness versus $f_{NL}$, with $\Delta_{\mathcal{R}}^2 = 2.46 \times 10^{-9}$ (its
WMAP5 value).  The dependence is not precisely captured by a simple scaling as for $\sigma_8$ (note that the axes are log-linear), but the curves are well-fit by cubic polynomials.}
  \end{center}
\end{figure}

As a second example, we compute the tSZ variance and skewness for a fixed WMAP5 background cosmology with a non-zero value of $f_{NL}$ added.  Again, to be clear, we fix $\Delta_{\mathcal{R}}^2 = 2.46 \times 10^{-9}$ (its WMAP5 value), not $\sigma_8 = 0.817$, as the non-zero value of $f_{NL}$ will change the value of $\sigma_8$ inferred at the low redshifts from which the tSZ signal originates.  We compute the change in the mass function due to primordial non-Gaussianity using the results of \cite{LoVerde-Smith2011}.  Although this is technically a ``friends-of-friends'' (FOF) mass function, we follow~\cite{Wagneretal2010} in assuming that the \emph{ratio} of the non-Gaussian mass function to the Gaussian mass function is nearly universal (and hence applicable to both FOF and spherical overdensity mass functions), even if the underlying mass function itself is not.  While this assumption may not be valid at the percent level, our approach should capture the relevant physical effects of primordial non-Gaussianity on the tSZ moments.  In particular, $f_{NL} > 0$ leads to a significant increase in the number of massive halos at late times, while $f_{NL} < 0$ has the opposite effect~\cite{Dalaletal2008,LoVerde-Smith2011}.  Thus, the amplitudes of the tSZ statistics should increase or decrease accordingly, as they are sourced by these massive halos.

The results of this analysis are shown in Figs.~10 and 11.  As expected, the tSZ variance and skewness both exhibit sensitivity to $f_{NL}$, although for the currently allowed range of values ($-10 < f_{NL} < 74$~\cite{Komatsuetal2011} for the ``local'' shape) it is overwhelmed by the uncertainty in the ICM astrophysics. However, using the method outlined in the previous section to constrain the ICM astrophysics and thus break its degeneracy with cosmology, a tight constraint on primordial non-Gaussianity with the tSZ effect might be achievable.  In addition, it is worth noting that the tSZ moments are sensitive to non-Gaussianity on cluster scales, rather than the large scales probed by the primordial CMB or large-scale halo bias.  Thus, the tSZ signal can constrain $f_{NL}$ on much smaller scales than other observables, allowing for complementary constraints that are relevant for models of inflation that predict strongly scale-dependent non-Gaussianity~\cite{LoVerdeetal2008}.

\subsection{Estimating the constraints from \emph{Planck}}
We briefly estimate the expected constraints on $\sigma_8$ and $\Sigma
m_{\nu}$ using tSZ measurements from \emph{Planck}.  \emph{Planck}'s
spectral coverage will allow the separation of the tSZ signal from
other CMB components, likely yielding a full-sky tSZ map.  However,
given issues with bandpass uncertainties and CO contamination, it is
difficult to predict the SNR for a \emph{Planck} measurement of
$\tilde{S}_{3,\beta=1.4}$.  Based on the Sehgal simulation analysis
above (which covers an octant of the sky), we anticipate that a cosmic
variance-limited full-sky measurement could achieve a SNR $\approx 90$
for the tSZ variance or a SNR $\approx 35$ for the tSZ skewness.  For
$\tilde{S}_{3,\beta=1.4}$, we estimate a SNR $\approx 55$, which
implies that the ICM astrophysics can be constrained to the $\lesssim
5$\% level.  This result will allow for a $\lesssim 1$\% constraint on
$\sigma_8$ after subsequently applying the correct model to interpret
tSZ measurements.  Note that these estimates fully account for both
Poisson and cosmic variance error, as they are derived from a
cosmological simulation.  Technically, we have neglected the fact that
the cosmic variance error will scale with $\sigma_8$ (see the
discussion in \cite{Wilsonetal2012}), but as long as the true value is
not so far from $\sigma_8 = 0.8$ as to conflict with all recent measurements of
this parameter, our estimates will be accurate.

Forecasting a $\Sigma m_{\nu}$ constraint is somewhat more difficult because it is highly degenerate with $\sigma_8$, and we have not accounted for other possible parameter dependences (e.g., one might expect $\Omega_m h^2$ to enter through its role in the mass function).  However, as described in~\cite{Ichiki-Takada2011}, %(and mentioned earlier in~\cite{Pierpaoli2003})
the effect of massive neutrinos is essentially captured by a low-redshift suppression of $\sigma_8$ as compared to the value inferred from the primordial CMB at $z = 1100$ (i.e., $\Delta_{\mathcal{R}}^2$).  This line of reasoning was also used in the neutrino mass constraint derived by the \emph{Chandra} Cluster Cosmology Project~\cite{Vikhlininetal2009}, who found an upper bound of $\Sigma m_{\nu} < 0.33$ eV at $95$\% CL.  Based on the results of~\cite{Ichiki-Takada2011}, $\Sigma m_{\nu} \approx 0.1$ eV is roughly equivalent to a $\approx 3$\% decrease in $\sigma_8$ as inferred at $z=1$ compared to $z=1100$.  The WMAP5 measurement of $\sigma_8$ has an uncertainty of $\approx 3$\%~\cite{Komatsuetal2009}, and the equivalent result from \emph{Planck} should be $\lesssim 1$\%~\cite{PlanckBlueBook}.  If the technique that we have outlined works as planned, it could yield a $\lesssim 1$\% measurement of $\sigma_8$ at low redshift. This corresponds to a sensitivity to $\Sigma m_{\nu} \approx 0.1$--$0.2$ eV.  Given the known lower bound of $\Sigma m_{\nu} \gsim 0.05$ eV, it might even be possible to detect the sum of the neutrino masses through tSZ measurements.  Clearly the \emph{Planck} results will have a smaller SNR than the full-sky, cosmic variance-limited assumption made above --- and there are important degeneracies between the relevant parameters --- but nonetheless these methods show great promise in constraining any parameter that affects the tSZ signal through the mass function.

\acknowledgments{We thank Nick Battaglia, Eiichiro Komatsu, Marilena LoVerde, Daisuke Nagai, Jeremiah Ostriker, Neelima Sehgal, Laurie Shaw, David Spergel, and Matias Zaldarriaga for useful conversations.  We also acknowledge Eiichiro Komatsu's online cosmology routine library, which was a helpful reference.  As this work neared completion, we became aware of related results from S.~Bhattacharya et al.~\cite{Bhattacharyaetal2012}, with whom we subsequently exchanged correspondence.  BDS was supported by an NSF Graduate Research Fellowship.}

\begin{figure}
\label{fig.sig8ratio}
  \begin{center}
    \includegraphics[width=\columnwidth]{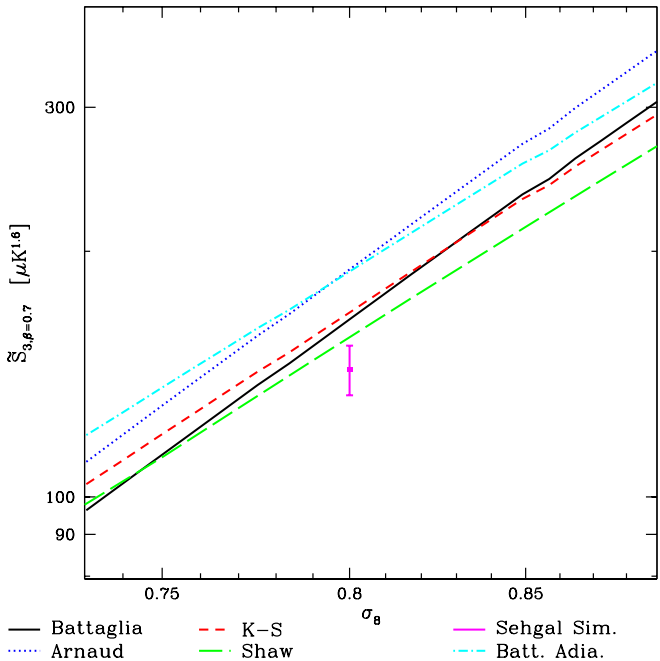}
    \caption{The rescaled skewness $\tilde{S}_{3,\beta}$ for $\beta=0.7$ plotted against $\sigma_8$.  Clearly there is significantly less scatter for this statistic between the different pressure profiles, but it still scales as $\sigma_8^{5-6}$.  Thus, this value of $\beta$ provides a method for constraining cosmological parameters that is fairly independent of the details of the gas physics.  The simulation point is somewhat low, but the disagreement is not statistically significant.  This point was also not used in determining the optimal value of $\beta$.}
  \end{center}
\end{figure}

\begin{figure}
\label{fig.masscontrib}
  \begin{center}
    \includegraphics[width=\columnwidth]{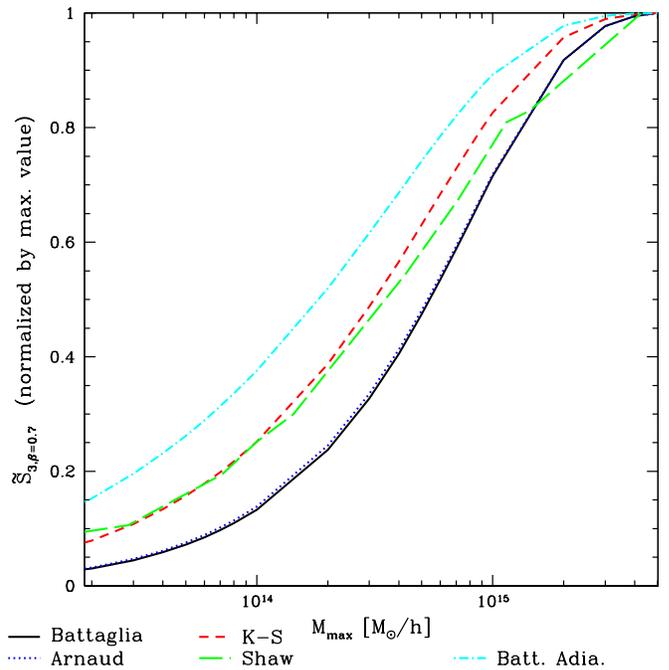}
    \caption{Contributions to the rescaled skewness with $\beta=0.7$ from clusters of various masses, for each of the different pressure profiles.  The contributions to this statistic are very similar to those found for the skewness in Fig.~4, as one would expect based on the definition of $\tilde{S}_{3,\beta=0.7}$.}
  \end{center}
\end{figure}

\appendix
\section{Directly Canceling the Gas Physics}
In this appendix, we present a different approach to circumventing the degeneracy between ICM astrophysics and cosmological parameters in tSZ measurements.  In particular, we investigate the rescaled skewness statistic in Eq.~(\ref{eq.ratiostat}) for other values of the exponent $\beta$ to see if they might directly cancel the gas physics dependence, but preserve a sensitivity to the underlying cosmology.  We heuristically motivate the magnitude of this exponent in the following way. We consider the effect of increasing the amount of energy input through feedback on both the tSZ variance and skewness in a theoretical model for the ICM. Increasing the feedback should substantially reduce the signal from low-mass clusters, and generally affect the pressure profile in the outer regions of clusters. However, high-mass clusters are much less affected by feedback, so their tSZ signal should be essentially unchanged. This fact implies that the variance, which depends more on the signal from low-mass clusters (as shown earlier) and receives contributions at large distances from the cluster center, should be reduced much more by feedback than the skewness, which depends more on the signal from massive clusters and is mostly produced in the inner regions of clusters. In order to construct a rescaled skewness in Eq.~(\ref{eq.ratiostat}) that is unchanged as the amount of astrophysical feedback is increased, we expect to exponentiate the variance (which is more sensitive to feedback) to a lower power than the skewness, in order to compensate.  Thus, we anticipate $\beta<1$ for a statistic that is somewhat insensitive to the ICM astrophysics, but remains sensitive to the underlying cosmology.

To derive this exponent quantitatively, we find the value of $\beta$ that minimizes the scatter between different choices of pressure profile for the same values of the cosmological parameters.  We focus specifically on $\sigma_8$, as the tSZ moments are most sensitive to this parameter.  For $0.73 < \sigma_8 < 0.9$ (and all other parameters taking their WMAP5 values), we find that $\beta = 0.7$ minimizes the scatter between the different pressure profiles.  As these profiles span a wide range of ICM prescriptions and tSZ power spectra predictions, we expect this result to be fairly robust.  We verify this claim by computing the rescaled skewness with $\beta=0.7$ for the Sehgal simulation.  The results are shown in Fig.~12.  Although the simulation point is somewhat low, it is only $1\sigma$ away from the halo model-based results at $\sigma_8=0.8$.

In order to gain more understanding of the rescaled skewness with $\beta=0.7$, we also investigate the characteristic mass scales that contribute to this statistic. The results are shown in Fig.~13. For a WMAP5 cosmology, we find that $\tilde{S}_{3,\beta=0.7}$ receives $\approx 25-50$\% of its amplitude from clusters with $M < 2$--$3 \times 10^{14} \,\, \mathrm{M}_{\odot}/h$.  In general, the signal for $\tilde{S}_{3,\beta=0.7}$ is dominated by mass scales similar to those that comprise the skewness, i.e., higher masses than those responsible for the variance.  It is interesting to note that this statistic receives contributions from rather different mass scales for the different profiles in Fig.~13, even though the final result is fairly independent of the profile choice (as seen in Fig.~12).

We also note that the same statistic with $\beta=0.7$ behaves similarly when applied to $\Sigma m_{\nu}$, as seen in Fig.~14.  To be clear, we do not re-fit for the value of $\beta$ that minimizes the difference between the various pressure profiles in this plot.  If we did perform such a fit,  the best-fitting exponent to minimize the dispersion is again $\beta=0.7$, in exact agreement with the best-fitting value for $\sigma_8$.

\begin{figure}
\label{fig.Mnuratio}
  \begin{center}
    \includegraphics[width=\columnwidth]{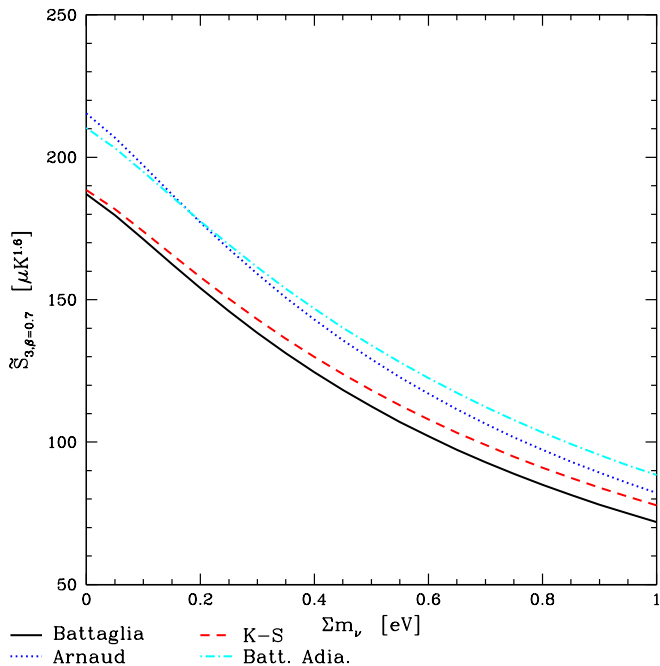}
    \caption{Similar to Fig.~12, but now showing the rescaled skewness with $\beta=0.7$ plotted against the sum of the neutrino masses $\Sigma m_{\nu}$. We assume $\Delta_{\mathcal{R}}^2 = 2.46 \times 10^{-9}$ (its WMAP5 value). There is significantly reduced scatter between the pressure profiles for this statistic, and it retains a leading-order quadratic dependence on $\Sigma m_{\nu}$.}
  \end{center}
\end{figure}

Despite the promising nature of the $\tilde{S}_{3,\beta=0.7}$ statistic, we must emphasize that we lack a strong theoretical motivation for this quantity, and thus one must be cautious in assuming its insensitivity to changes in the gas physics.  For example, if one alters the amount of non-thermal pressure support by changing an overall parameter that affects the normalization of all halos' pressure profiles (e.g., the $\alpha_0$ parameter in the Shaw model~\cite{Shawetal2010}), then this statistic will be affected in a nontrivial manner.  It is likely that requiring the model to match the observed X-ray data for massive, low-redshift clusters will prevent a drastic change along these lines, but it is still a possibility.  In addition, as mentioned earlier, the tSZ skewness is fairly sensitive to the choice of mass function used in these calculations (though the variance is less affected), which will add additional scatter and uncertainty to the relation presented in Fig.~12.

Nonetheless, we are encouraged by the mild sensitivity of $\tilde{S}_{3,\beta=0.7}$ to the complicated astrophysics of the ICM.  Moreover, it retains a strong dependence on $\sigma_8$.  In particular, using the results presented in Table~I for $\langle T^2 \rangle$ and $\langle T^3 \rangle$, we find that $\tilde{S}_{3,\beta=0.7} \propto \sigma_8^{5-6}$.  For a fixed value of $\sigma_8=0.817$, the scatter in $\tilde{S}_{3,\beta=0.7}$ between the different profiles is only $7$\%.  This scatter is much smaller than that in the variance or skewness: for the same value of $\sigma_8 = 0.817$, the scatter in $\langle T^2 \rangle$ is $\approx 50$\% between these profiles, while for $\langle T^3 \rangle$ it is $\approx 35$\%.  This statistic could thus be used for a high-precision determination of $\sigma_8$ from tSZ measurements, with significantly reduced sensitivity to theoretical systematic uncertainties due to unknown physics in the ICM.  For a full-sky, CV-limited experiment, we estimate a SNR $\approx 40$ for $\tilde{S}_{3,\beta=0.7}$, which implies a $\lesssim 1$\% constraint on $\sigma_8$, in agreement with the estimates presented for $\tilde{S}_{3,\beta=1.4}$ earlier.

However, before applying this technique to real data, its insensitivity to the gas physics prescription must be tested with detailed numerical tSZ simulations, as we lack a good understanding of the theoretical origin of $\tilde{S}_{3,\beta=0.7}$.  The rescaled skewness with $\beta=1.4$, though, is well-motivated theoretically and provides a clear method to constrain the ICM astrophysics in low-mass halos, which will greatly reduce the theoretical systematic uncertainty in tSZ measurements.  We defer an application of the $\beta=0.7$ approach to future work.

\begin{figure}
\label{fig.sig8ratiocancel2pt4pt}
  \begin{center}
    \includegraphics[width=\columnwidth]{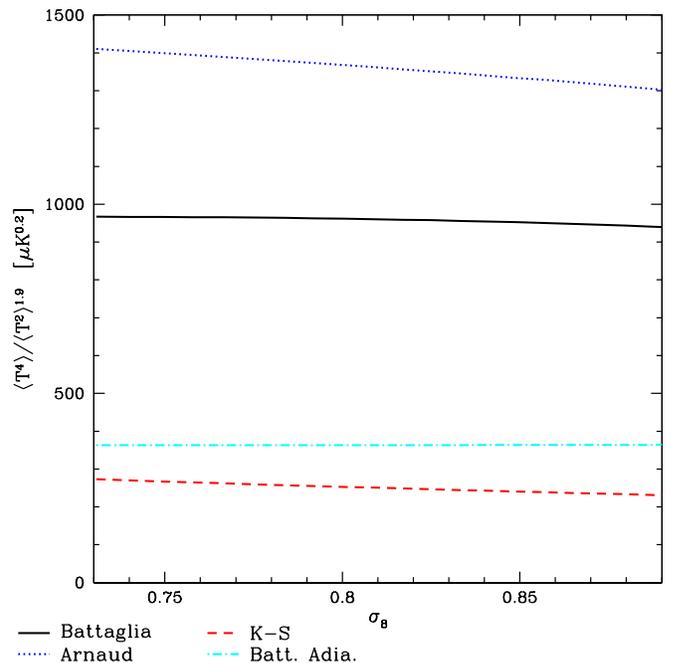}
    \caption{The rescaled kurtosis $\langle T^4 \rangle/\langle T^2 \rangle^{1.9}$ plotted against $\sigma_8$.  It is evident that this statistic is nearly independent of $\sigma_8$, as expected based on the scalings in Tables~I and~II.  However, it is still dependent on the ICM astrophysics, as represented by the different pressure profiles.  Thus, this statistic --- like the rescaled skewness --- can be used for determining the correct gas physics model.}
  \end{center}
\end{figure}

\section{Higher-Order tSZ Moments}
In the main text of the paper, we focus on the second (variance) and third (unnormalized skewness) moments of the tSZ signal.  In this appendix, we provide additional calculations of the fourth moment (the unnormalized kurtosis) and investigate a ``rescaled kurtosis'' statistic similar to the rescaled skewness defined in Eq.~(\ref{eq.ratiostat}).  The unnormalized kurtosis is simply given by Eq.~(\ref{eq.Npoint}) with $N=4$.  As for the variance and skewness, we compute this quantity for each of the various pressure profiles described in the text, and investigate the dependence on $\sigma_8$.  We define an amplitude $A_4$ and scaling $\alpha_4$ precisely analogous to $A_{2,3}$ and $\alpha_{2,3}$ in Eq.~(\ref{eq.powerlaw}).  The results for each of the profiles are given in Table~II.

\begin{table}
\begin{tabular}{c | c | c}
\label{tab.scalings2}
Profile & $A_4$ [$10^5$ $\mu$K$^4$] & $\alpha_4$ \\
\hline
Arnaud & $4.22$ & $14.7$ \\
Battaglia & $3.58$ & $14.5$ \\
Batt.\ Adiabatic & $5.48$ & $12.5$ \\
Komatsu-Seljak & $4.70$ & $13.5$ \\
\end{tabular}
\caption{Amplitudes and power-law scalings with $\sigma_8$ for the tSZ kurtosis. The first column lists the pressure profile used in the calculation (note that all calculations use the Tinker mass function).  The amplitudes are specified at $\sigma_8 = 0.817$, the WMAP5 maximum-likelihood value.  All results are computed at $\nu = 150$ GHz.}
\end{table}

As for the variance and skewness, the profiles that account for feedback processes (Battaglia and Arnaud) have a steeper scaling with $\sigma_8$ than those that do not, because the kurtosis signal for these profiles is dominated to an even greater extent by the most massive, rare halos.  Additionally, note that the scatter between the various profiles is even smaller than that for the skewness (and much smaller than that for the variance).

\begin{figure}
\label{fig.2pt4ptsig8ratiocancelvsMmax}
  \begin{center}
    \includegraphics[width=\columnwidth]{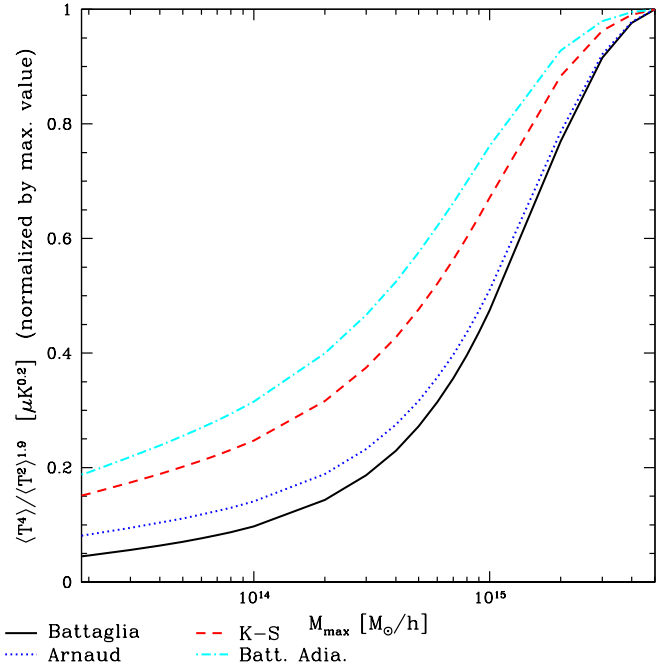}
    \caption{Fraction of the rescaled kurtosis $\langle T^4 \rangle/\langle T^2 \rangle^{1.9}$ contributed by clusters with virial mass $M < M_{\mathrm{max}}$.  Although a non-trivial fraction of the signal comes from low-mass objects, comparison with Fig.~6 indicates that the rescaled kurtosis is mostly sourced by more massive halos than those responsible for the rescaled skewness.}
  \end{center}
\end{figure}

Furthermore, it is possible to construct a ``rescaled kurtosis'' statistic similar to the rescaled skewness defined in Eq.~(\ref{eq.ratiostat}), which should be nearly independent of the background cosmology.  In particular, based on the scalings in Tables~I and~II, one can immediately see that the quantity $\langle T^4 \rangle/\langle T^2 \rangle^{1.9}$ effectively cancels the dependence on $\sigma_8$ for each of the profiles.  This result is clearly seen in Fig.~15, which presents this statistic as a function of $\sigma_8$.  Encouragingly, a clear dependence on the ICM astrophysics persists, as for the rescaled skewness in Fig.~5.  Note that the same hierarchy is seen between the different profiles in both figures --- again, this arises because of the suppression of the tSZ signal in low-mass halos for the profiles that include significant feedback, which decreases their value of the variance, thus increasing their value of the rescaled skewness and kurtosis.  In addition, it appears that the rescaled kurtosis (Fig.~15) is somewhat less dependent on $\sigma_8$ than the rescaled skewness (Fig.~5), although the difference is not very significant over the feasible range of $\sigma_8$.  Physically, this is likely due to the fact that the kurtosis is dominated even more than the skewness by very massive halos, for which the different pressure profiles do not differ widely in their predictions.  Thus, the various profiles have a similar relative scaling between the kurtosis and the variance, making the rescaled kurtosis nearly independent of $\sigma_8$ for the same choice of ``rescaling exponent''.  We verify this interpretation in Fig.~16, which shows the fraction of this signal contributed by clusters with virial mass $M < M_{\mathrm{max}}$.  Compared to the rescaled skewness (Fig.~6), the rescaled kurtosis is sourced more predominantly by fairly massive halos.  The rescaled kurtosis thus also provides a route to determining the gas physics of the ICM, nearly independent of the background cosmology.

Unfortunately, a measurement of the tSZ kurtosis appears extremely challenging with current data, both because of confusion with other signals (e.g., the kinetic SZ effect, point sources, and so on) and because the cosmic variance for this statistic is quite large, as it is sourced by very massive clusters.  Previous work on the kinetic SZ kurtosis allows some rough estimates of this contamination to be made, however~\cite{Zhangetal2002,Castro2004}.  In~\cite{Zhangetal2002}, estimates are given of the normalized (dimensionless) skewness and kurtosis for the tSZ and kSZ effects, as determined from hydrodynamical simulations.  Based on their results, it appears that the dimensionless kurtosis due to the tSZ effect is at least a few times larger than that due to the kSZ effect.  Thus, assuming that the variance of the kSZ signal is no larger than that of the tSZ signal, this implies that the value of $\langle T^4 \rangle$ that we have calculated for the tSZ effect will be at least a few times larger than the contribution due to the kSZ effect (at 150 GHz).  Hence, the kSZ contamination may not be a major problem for a detection of the tSZ kurtosis.  However, contamination from point sources will likely remain a significant problem for an experiment with only one or two frequencies.  Thus, ACT and SPT are unlikely to make a detection, though \emph{Planck} may do so.  Neglecting any potential contaminants and using the Sehgal simulation analysis described earlier, we find that a cosmic variance-limited full-sky experiment could achieve a SNR $\approx 18$ for the tSZ kurtosis or a SNR $\approx 22$ for the rescaled kurtosis.  The results from \emph{Planck} will clearly have a somewhat lower SNR than these estimates, but it may still be worthwhile to investigate these quantities using the forthcoming \emph{Planck} sky maps.

%\bibliography{Hill-Sherwin}

\end{document}